\documentclass[aps,twocolumn,showpacs,prx,amsmath,amssymb,floatfix,superscriptaddress]{revtex4-1}

\usepackage{color}
\usepackage{bbm}
\usepackage{graphicx}
\usepackage{dcolumn}
\usepackage{bm}% bold math
\usepackage{array}
\usepackage{float}
\usepackage{supertabular}
\usepackage{longtable}
\usepackage{mathrsfs}
\usepackage{txfonts}
\usepackage[bbgreekl]{mathbbol}
\usepackage{wasysym}
\usepackage[T1]{fontenc}
\usepackage[usenames,dvipsnames]{xcolor}
\usepackage{amsmath}
\usepackage[colorlinks=true,citecolor=Cerulean,linkcolor=RubineRed,urlcolor=Cerulean]{hyperref}

\begin{document}

\title{Engineering quasi-steady-state correlations in uncorrelated thermal states using stochastic driving}

\author{Armin Rahmani}
\affiliation{Department of Physics and Astronomy and Advanced Materials Science and Engineering Center, Western Washington University, Bellingham, Washington 98225, USA}
\affiliation{Kavli Institute for Theoretical Physics, University of California, Santa Barbara, California 93106, USA}

\date{\today}

\begin{abstract}
Nonequilibrium quantum dynamics can give rise to the emergence of novel steady states. We propose a scheme for driving an initially uncorrelated thermal state to generate customized correlation functions by determining and reverse engineering the steady-state two-point functions for a class of Markov processes. We also extend the formalism to the calculation of four-point functions. We then apply our method to generating power-law correlated fermionic Green's functions. Furthermore, we find that the power-law patterns emerge at much shorter times than the convergence to the steady state, at which point the disorder in the two-point correlations disappears. On the other hand, the density-density correlations exhibit steady-state disorder while following a power-law trendline. These ideal steady states appear as intermediate-time quasi-steady states in the presence of perturbations.

\end{abstract}

\maketitle
\section{Introduction }
There has been significant progress in driving nonequilibrium quantum matter to prepare desired quantum states, particularly in the context of quantum simulations~\cite{Georgescu14,Biamonte11}. While the objective is often creating states with equilibrium counterparts, such as the ground state of interacting Hamiltonians~\cite{OC1,OCrev2}, it has been realized that quantum driving may give rise to the emergence of states without equilibrium counterparts. Floquet engineering, i.e., controlling the evolution by periodic driving, has been a fertile ground for generating such states~\cite{PhysRevX.4.031027,Oka,
Qin}. Furthermore, dissipation engineering has been similarly effective for open systems~\cite{Zoller2008,Diehl2008,Busiello_2018,Marino2022,Seetharam2022,Seetharam2022_2}.

In this paper, we focus on stochastic drives~\cite{Strunz,Silva2012,Pichler2013,Luca2013,Chenu} corresponding to pure dephasing for noninteracting fermions \cite{Trajectories}, with formal similarities to open systems~\cite{Prosen_2008,_nidari__2010,Eisler_2011,Marko2}. We propose a scheme for creating almost arbitrary two-point correlation functions starting from uncorrelated thermal states. Our method utilizes a single temporally uncorrelated noise source coupled to a fixed drive Hamiltonian, resulting in spatially correlated noise. The Lindblad master equation governing the dynamics only contains one double commutator, analogous to a fully dissipative open system.

 We show that, with this scheme, we can engineer a broad class of custom-ordered equal-time two-point Green's functions (also known as the single-particle density matrix~\cite{Eckardt2018}). Long-range couplings~\cite{Britton2012,Richerme2014,Jurcevic2014} are necessary for creating long-range power-law correlations in the shortest possible time for a drive Hamiltonian with a fixed norm (See Ref.~\cite{Barthel} for general results on limitations of creating critical correlations in open systems). As two-point correlations do not fully characterize a quantum system's many-body wave function, we also calculate the long-time limit of the four-point density-density correlation function to further investigate the nature of the steady states. We find that while the steady-state two-point function decays as a clean power law, the density-density correlator remains highly disordered but follows a power-law trendline with a different exponent.
 
Under ideal conditions, our pure dephasing dynamics allow the system to escape the generic fate of a noise-driven quantum state; namely, a featureless infinite-temperature fixed point originating from a noise-induced heating rate that constantly deposits energy into the system~\cite{Bunin2011,Luca2013,Rahmani2014}. This scheme relies on the predominance of a single noise source, engineered using high-frequency pulse control. To achieve pure dephasing, we need to eliminate other noise sources or nonstochastic contributions to the Hamiltonian. In practice, however, even if we engineer our spatially correlated noise to have the largest energy scale, perturbations will ultimately force the system to flow to the infinite-temperature fixed point. However, for infinitesimally small perturbations (compared with the energy scale of the engineered noise, which can be made as large as possible within the experimental constraints), the timescales associated with the engineered noise and the perturbations separate. Thus, the steady states predicted in this paper for the ideal case are expected to emerge as slightly perturbed quasi-steady states at intermediate timescales before the system heats up to infinite temperature at much longer evolution times.

A noise term dominating the dynamics can only be generated in a controllable synthetic platform, with a strong random signal driving the system and naturally occurring noise sources reduced as much as possible. Two experimental platforms are potential candidates for our proposed correlation engineering method: cold fermionic atoms in optical lattices and digitized quantum processors with pulse level control. For a short-range drive Hamiltonian, fermionic atoms are the most convenient but creating direct long-range couplings is challenging with cold atoms. However, there has been recent progress in simulating long-range hopping with temporal driving~\cite{Martinez2021,Roses2021,Bastidas2022}. On the other hand, general-purpose digitized devices can incorporate nonlocal couplings (either directly designed into the hardware or by using swap gates). Although qubits in these devices are not fermionic, the fermionic exchange statistics can be implemented through Jordan-Wigner or Braviy-Kitaev mappings~\cite{Abrams1997,Bravyi2002,Somma2002,Verstraete2005,Whitfield2011}.

The outline of this paper is as follows. In Sec.~\ref{sec:formalism}, we extend the results of Ref.~\cite{Rahmani2014} on the steady-state two-point functions to steady-state four-point functions under stochastic dephasing dynamics. In Sec. \ref{sec:green}, we present our method for reverse engineering the drive to provide the desired matrix of fermionic equal-time Green's functions. In Sec.~\ref{sec:power}, we apply our method to the case of power-law correlated Green's functions. In Sec. \ref{sec:NN}, we examine the steady-state density-density correlation functions for the drive used in engineering the Green's function. Finally, we present our conclusions in Sec. \ref{sec:conclusions}.

\section{Steady-state two- and four-point functions for fully stochastic quadratic driving  \label{sec:formalism}}
Here we present the theoretical framework for stochastic dynamics capable of generating customized noise-averaged correlations in fermionic systems, initially in an uncorrelated thermal state. For a system of $N$ spinless fermions, where the creation/annihilation operators satisfy the anticommutation relation  $\{c^\dagger_i, c_j\}=\delta_{ij}$, consider a quadratic Hamiltonian that stochastically fluctuates around zero. We write $H(t)$ as
\begin{equation}\label{eq:drive}
H(t)=\xi(t)V, \quad V=\sum_{ij}J_{ij}c^\dagger_i c_j, \quad t>0.
\end{equation}
Here, $J$ is an $N\times N$ Hermitian matrix, and $\xi(t)$ is Gaussian white noise with $\overline {\xi(t)}=0$ and $\overline{ \xi(t) \xi(t')}=w^2\delta(t-t')$. The overline symbol represents averaging over the realizations of noise. The noise strength $w$ sets the equilibration timescale associated with this noise term, but does not affect the nature of the steady state in the absence of perturbations. Although the couplings $J_{ij}$ are general, fixing the geometry is necessary for engineering correlations decaying as a power law of the distance between sites. We assume the sites as arranged on a one-dimensional line.

The evolution by $H(t)$ describes an isolated system with a vanishing average Hamiltonian driven by temporally uncorrelated and spatially correlated noise, with $J$ encoding the spatial correlations. The system will reach a steady-state density matrix in the long-time limit $t\to\infty$, given by
 \begin{equation}
 \rho(t\to\infty)=\lim_{t\to\infty} \overline {{\rm tr}\left[e^{-i\int_0^tH(t') dt'}\rho(0)e^{i\int_0^tH(t') dt'}\right]},
\end{equation}
where $\rho(0)$ is the initial density matrix.
Since the Hamiltonians at different times commute, time ordering is not necessary. We can view the dynamics as a Markov process described by a stochastic Schr\"odinger equation. The long-time limit of the dynamics can be alternatively viewed as a time average over random evolution times after a quench to Hamiltonian $V$, as the white noise is simply creating a random walk for the total evolution time with $V$.

The equation of motion for the density matrix is then a Lindblad master equation
$\partial_t \rho(t)=-{1\over 2}w^2[[\rho,V], V]$.
Similar to Ref.~\cite{Rahmani2014}, we write a Heisenberg-picture analog of the master equation for a noise-averaged Heisenberg operator $O(t)$ as
\begin{equation}\label{eq:master}
\partial_t O(t)=-{1\over 2}w^2[[O(t),V], V].
\end{equation}
Our analysis is focused on the unperturbed evolution above. However, in general we can have both stochastic and nonstochastic perturbations to $H(t)$, which give rise to $\partial_t O(t)=-{1\over 2}w^2[[O(t),V], V]+\epsilon i[H_1,O(t)]+\sum_j\epsilon_j^2[[O(t),V_j], V_j]$,
where $H_1$ is nonstochastic and $\epsilon_j$ are the strengths of other noise terms. These perturbations introduce a longer timescale than the timescale associated with $w$, in which the system generically flows to the infinite-temperature fixed point with a density matrix proportional to the identity. In the remainder of this paper, we assume our timescales are shorter than the time needed to reach infinite temperature, neglect the perturbations, and analyze the problem using the ideal master equation~\eqref{eq:master}. We also set $w$ to 1, which amounts to a simple rescaling of evolution time.

An important observation is that under evolution with Eq.~\eqref{eq:master}, generic quadratic (quartic) operators remain quadratic (quartic). In contrast, the evolution can spread operators into nontrivial linear combinations of other quadratic (quartic) operators.
%
%
% for the particular operator $c^\dagger_ic_j$, the Heisenberg operator $O^{ij}(t)$, which solves the master equation \eqref{eq:master} with initial condition $O^{ij}(0)= c^\dagger_ic_j$ retains the following form at all times:
%\begin{equation}\label{eq:ansatz}
%O^{ij}(t)=\sum_{\alpha\beta}{\mathscr O}^{ij}_{\alpha \beta}(t)c^\dagger_\alpha c_\beta,
%\end{equation}
%where ${\mathscr O}^{ij}(t)$ is a time-dependent $N\times N$ matrix. 
 The above observation follows from the fact that in the commutator of two-quadratic forms, the quartic terms cancel out and we are left with another quadratic form
$
\left[\sum_{\alpha \beta}A_{\alpha\beta}c^\dagger_\alpha c_\beta, \sum_{ij}B_{ij}c^\dagger_ic_j\right]=
\sum_{\alpha \beta}[A,B]_{\alpha\beta}c^\dagger_\alpha c_\beta$ \cite{Klich}.

 We now consider a general quadratic operator 
\begin{equation}
O(t)=\sum_{\alpha\beta}{\mathscr O}_{\alpha \beta}(t)c^\dagger_\alpha c_\beta,
\end{equation}
where $\mathscr O$ is an $N\times N$ matrix.
 In terms of the $N\times N$ matrix $J$ [see Eq.~\eqref{eq:drive}], we can then write the solution of the master equation~\eqref{eq:master} as \cite{Rahmani2014}
\begin{equation}\label{eq:master_sol}
{\mathscr O}_{\alpha \beta}(t)=\sum_{\sigma \lambda \eta \gamma}{\mathscr U}_{\alpha \sigma} {\mathscr U}^\dagger_{ \lambda \beta}
e^{-(1/ 2)\left({\cal D}_{ \sigma }-{\cal D}_{ \lambda}\right)^2 t}
{\mathscr U}^\dagger_{ \sigma \eta}{\mathscr U}_{ \gamma \lambda}  {\mathscr O}_{\eta \gamma}(0),
\end{equation}
where column $i$ of the unitary matrix $\mathscr U$ is the eigenvector of $J$ with eigenvalue ${\cal D}_i$. In other words,  
\begin{equation}\label{eq:J_diag}
J= {\mathscr U}{\rm diag}({\cal D}_1, {\cal D}_2, \dots ,{\cal D}_N){\mathscr U}^\dagger.
\end{equation}
We note that although all derivations in the paper are for white noise, the nature of the long-time limit also applies to a wider class of noise spectra with finite zero-frequency spectral power, e.g., the Ornstein-Uhlenbeck process effectively only rescales the evolution time compared to white noise~\cite{Luca2013}.

Assuming $J$ has a nondegenerate spectrum, we can then write the long-time limit of the noise-averaged Heisenberg operator for $t\to\infty$ as
\begin{equation}\label{eq:master_sol2}
{\mathscr O}_{\alpha \beta}(\infty)=\sum_{  \eta \gamma}\left(\sum_\lambda{\mathscr U}_{\alpha \lambda} {\mathscr U}^\dagger_{ \lambda \beta}
{\mathscr U}^\dagger_{ \lambda \eta}{\mathscr U}_{ \gamma \lambda} \right) {\mathscr O}_{\eta \gamma}(0).
\end{equation}

We now extend this formalism to quartic operators. Of particular interest are density-density correlation functions. To find the steady-state correlation functions, again we use the Heisenberg-picture approach. Due to the cancellation of the sextic terms in the commutator of a quartic form and a quadratic from $V=\sum_{ij}J_{ij}c^\dagger_i c_j$, we find
\begin{equation}
\left[\sum_{\alpha\beta\gamma\delta}{\mathscr R}_{\alpha\beta\gamma\delta}(t)c^\dagger_\alpha c_\beta c^\dagger_\gamma c_\delta, V\right]=\sum_{\alpha\beta\gamma\delta}{\mathscr S}_{\alpha\beta\gamma\delta}(t)c^\dagger_\alpha c_\beta c^\dagger_\gamma c_\delta,
\end{equation}
where \begin{equation}\label{eq:map}
{\mathscr S}_{\alpha\beta\gamma\delta}=\sum_j\left({\mathscr R}_{\alpha\beta\gamma j}J_{j\delta}-{\mathscr R}_{\alpha\beta j \delta}J_{\gamma j}+{\mathscr R}_{\alpha j\gamma\delta}J_{j\beta}-{\mathscr R}_{j\beta\gamma\delta}J_{\alpha j}\right).
\end{equation}
Therefore, all quartic operators of the form above retain this structure
%
%
%
%
%
%a quadratic operator, which is initially equal to a product of two density operators $R^{ij}(0)=c^\dagger_ic_ic^\dagger_jc_j$, retains the form
%\begin{equation}
%R^{ij}(t)=\sum_{\alpha\beta\gamma\delta}{\mathscr R}_{\alpha\beta\gamma\delta}(t)c^\dagger_\alpha c_\beta c^\dagger_\gamma c_\delta,
%\end{equation} 
when evolving with the master equation~\eqref{eq:master}. By using Eq.~\eqref{eq:map} twice, we can write the equation of motion for the tensor $\mathscr R$ as 
\begin{equation}
\partial_t{\mathscr R}_{\alpha\beta\gamma\delta}=-{1\over 2}\sum_{xyzw}\big(Q_{\alpha  \beta\gamma \delta}^{ xy zw}+Q_{ \beta\gamma \delta\alpha }^{y zw x}+Q_{\gamma \delta\alpha  \beta}^{ zwxy }+Q_{ \delta\alpha  \beta\gamma}^{ wxy z}\big){\mathscr R}_{xyzw},
\end{equation}
%\begin{eqnarray}
%\partial_t{\mathscr R}_{\alpha\beta\gamma\delta}&=&\sum_{xyzw} {\cal L}^{xyzw}_{\alpha\beta\gamma\delta}{\mathscr R}_{xyzw},\\
%{\cal L}^{xyzw}_{\alpha\beta\gamma\delta}&=&-{1\over 2}\big(Q_{\alpha  \beta\gamma \delta}^{ xy zw}+Q_{ \beta\gamma \delta\alpha }^{y zw x}+Q_{\gamma \delta\alpha  \beta}^{ zwxy }+Q_{ \delta\alpha  \beta\gamma}^{ wxy z}\big),
%\end{eqnarray}
where 
%\begin{equation}
%\hat {\cal L}=-{1\over 2}\left(Q_{ABCD}+Q_{BCDA}+Q_{CDAB}+Q_{DABC}\right)
%\end{equation}
%\begin{equation}
%\begin{split}
%\hat {\cal L}^{xyzw}_{\alpha\beta\gamma\delta}=-{1\over 2}\big(Q_{\alpha x, y\beta,\gamma z,w\delta}+Q_{y\beta,\gamma z,w\delta,\alpha x}+Q_{\gamma z,w\delta,\alpha x, y\beta }+Q_{w\delta,\alpha x, y\beta,\gamma z}\big)
%\end{split}
%\end{equation}
\begin{equation}
\begin{split}
 Q_{\alpha  \beta\gamma \delta}^{ xy zw}=
 &-J_{\alpha x}J_{y\beta}\delta_{\gamma z}\delta_{w\delta}+J_{\alpha x}\delta_{y\beta}J_{\gamma z}\delta_{w\delta}-J_{\alpha x}\delta_{y\beta}\delta_{\gamma z}J_{w\delta}\\
 &+J^2_{\alpha x}\delta_{y\beta}\delta_{\gamma z}\delta_{w\delta}.
 \end{split}
\end{equation}
%with $(A,B,C,D)\equiv (\alpha x, y\beta,\gamma z,w\delta)$ and
%\begin{equation}
%Q_{ABCD}\equiv J^2_{A}\delta_B\delta_C\delta_D-J_{A}J_B\delta_C\delta_D+J_{A}\delta_BJ_C\delta_D-J_{A}\delta_B\delta_CJ_D,
%\end{equation}
%where $J^2_{\tiny A}\equiv J^2_{\alpha x}$ should be interpreted as the element $\alpha x$ of the matrix $J^2$ obtained by matrix multiplication.

We now use the diagonalization of $J$ [Eq.~\eqref{eq:J_diag}] to find an exact solution to the equation above. We write the Kronecker deltas as elements of ${\mathscr U}^\dagger{\mathscr U}$ to obtain
\begin{widetext}
\begin{equation}
{\mathscr R}_{\alpha\beta\gamma\delta}(t)=\sum_{xyzw,abcd}{\mathscr U}_{\alpha a}{\mathscr U}_{yb}{\mathscr U}_{\gamma c}{\mathscr U}_{wd}e^{-(1/2)\left({\cal D}_a+{\cal D}_c-{\cal D}_b-{\cal D}_d\right)^2t}{\mathscr U}^\dagger_{d\delta}{\mathscr U}^\dagger_{cz}{\mathscr U}^\dagger_{b\beta}{\mathscr U}^\dagger_{ax}{\mathscr R}_{xyzw}(0).
\end{equation}
In the absence of any accidental vanishing of the exponent, we can calculate the long-time limit of the $\mathscr R$ tensor by noting that we must have either $b=a$ and $d=c$ or $b=c$ and $d=a$ (which double counts the case with $a=b=c=d$)~\cite{Ziraldo,Rahmani2018}. Thus we find
% ${\mathscr R}_{\alpha\beta\gamma\delta}(t\to\infty)=\sum_{xyzw}{\cal G}^{xyzw}_{\alpha\beta\gamma\delta}{\mathscr R}_{xyzw}(0)$ with
\begin{equation}\label{eq:4_long}
{\mathscr R}_{\alpha\beta\gamma\delta}(\infty)=\sum_{xyzw}\left[\sum_{ac}
{\mathscr U}_{\alpha a}{\mathscr U}_{\gamma c}{\mathscr U}^\dagger_{cz}{\mathscr U}^\dagger_{ax}
\left({\mathscr U}_{ya}{\mathscr U}_{wc}{\mathscr U}^\dagger_{c\delta}{\mathscr U}^\dagger_{a\beta}+{\mathscr U}_{yc}{\mathscr U}_{wa}{\mathscr U}^\dagger_{a\delta}{\mathscr U}^\dagger_{c\beta}\right)-
\sum_a{\mathscr U}_{\alpha a}{\mathscr U}_{\gamma a}{\mathscr U}^\dagger_{az}{\mathscr U}^\dagger_{ax}
{\mathscr U}_{ya}{\mathscr U}_{wa}{\mathscr U}^\dagger_{a\delta}{\mathscr U}^\dagger_{a\beta}\right]{\mathscr R}_{xyzw}(0).
\end{equation}
\end{widetext}
For the density-density correlator $n_1 n_j=c^\dagger_i c_i c^\dagger_j c_j$, we have
\[
{\mathscr R}_{xyzw}(0)=\delta_{xi}\delta_{yi}\delta_{zj}\delta_{wj}.
\]
However, we need to work with the more general tensors above as they appear in the Heisenberg-picture operator.

\section{Engineering steady-state Green's functions in uncorrelated thermal states\label{sec:green}}
An important class of correlation functions, i.e., the equal-time single-particle Green's functions, are given by the expectation values
\begin{equation}\label{eq:corr}
G_{ij}(t)=\langle c^\dagger_i c_j\rangle={\rm tr}\left[\rho (t) c^\dagger_i c_j \right],
\end{equation}
where $\rho(t)$ is the density matrix of the system at time $t$.
We can collect these correlators into an $N\times N$ Hermitian matrix $G$, with elements $G_{ij}=G^*_{ji}$. 

Let us consider a system initially prepared in a trivial thermal state at inverse temperature $\bbbeta$ with the density matrix
\begin{equation}
\rho(0)=\exp\left(-\bbbeta H_0\right)/Z, \quad H_0=\sum_i \mu_i\hat n_i,
\end{equation}
 with the partition function $Z={\rm tr}[\exp\left(-\bbbeta H_0\right)]$.
As $H_0$ is diagonal in the occupation-number basis, the different sites are uncorrelated in the thermal state above and we can write the initial Green's function as
% $G_{ij}^0={1\over Z}{\sum_{\{n_k=0,1\}}\langle n_1\dots n_N|c^\dagger_ic_j|n_1\dots n_N\rangle e^{-\bbbeta(\mu_1n_1+\dots \mu_Nn_N)}  }$, which leads to
\begin{equation}\label{eq:corr0}
G_{ij}(0)=f(\mu_j)\delta_{ij}={1\over 1+e^{\bbbeta \mu_i}}\delta_{ij},
\end{equation}
where $f$ represents the Fermi-Dirac function.

Suppose we drive this system by a stochastic quadratic Hamiltonian as in Eq.~\eqref{eq:drive}. Can we engineer custom-ordered Green's functions in the steady state of the system that emerges in the long-time limit?
%
%
%The occupation-number basis $|n_1n_2\dots n_N\rangle=(c^\dagger_1)^{n_1}(c^\dagger_2)^{n_2}\dots(c^\dagger_N)^{n_N}
%|0\rangle$, with $n_i=0,1$, is a natural choice for representing the states in the Hilbert space. Here $|0\rangle$ is he vacuum and $(c^\dagger_i)^0$ is defined as the identity operator.
%
More precisely, suppose our goal is choosing $\mu_i$ and $J_{ij}$ such that 
\begin{equation}
\lim_{t\to\infty}\overline{{\rm tr}\left[U(t)\rho_0U^\dagger(t)c^\dagger_ic_j\right]}=G_{ij},
\end{equation}
for \textit{given} (perhaps even arbitrary) Hermitian correlation matrix $G$ and initial inverse temperature $\bbbeta$.

For the $c^\dagger_i c_j$ operator, we have ${\mathscr O}_{\eta \gamma}(0)=\delta_{\eta i}\delta_{\gamma j}$, which leads to the following matrix elements when inserted in Eq.~\eqref{eq:master_sol}:
\begin{equation}
{\mathscr O}_{\alpha \beta}(t)=\sum_{\sigma \lambda }{\mathscr U}_{\alpha \sigma} {\mathscr U}^\dagger_{ \lambda \beta}
e^{-(1/2)\left({\cal D}_{ \sigma }-{\cal D}_{ \lambda}\right)^2 t}
{\mathscr U}^\dagger_{ \sigma i}{\mathscr U}_{ j \lambda}  
\end{equation}
Since we are engineering the matrix $J$, we can require it to have a nondegenerate spectrum. Thus, analogously to Eq.~\eqref{eq:master_sol2}, the Heisenberg-picture matrix corresponding to the long-time limit of the $c^\dagger_i c_j$ operator is given by 
\begin{equation}
{\mathscr O}_{\alpha \beta}(\infty)=\sum_\lambda{\mathscr U}_{\alpha \lambda} {\mathscr U}^\dagger_{ \lambda \beta}
{\mathscr U}^\dagger_{ \lambda i}{\mathscr U}_{ j \lambda}
\end{equation}
The above expressions lead to the relationships below between the time-dependent and steady-state correlation functions and the initial correlation functions through the eigenvectors of $J$:
\begin{eqnarray}\label{eq:G}
G_{ij}(t)&=&\sum_{\alpha \beta \lambda \sigma}{\mathscr U}_{j\lambda}{\mathscr U}^\dagger_{\lambda \beta}G_{\alpha\beta}(0)e^{-(1/2)\left({\cal D}_{ \sigma }-{\cal D}_{ \lambda}\right)^2 t}{\mathscr U}_{\alpha\sigma}{\mathscr U}^\dagger_{\sigma i},\\
G_{ij}&=&G_{ij}(\infty)=\sum_{\lambda}{\mathscr U}_{j\lambda}\left(\sum_{\alpha\beta}{\mathscr U}^\dagger_{\lambda \beta}G_{\alpha\beta}(0){\mathscr U}_{\alpha\lambda}\right){\mathscr U}^\dagger_{\lambda i}.
\end{eqnarray}

 The above equation implies that $G^T={\mathscr U}{\mathscr A}{\mathscr U}^\dagger$, where ${\mathscr A}$ is a diagonal matrix with 
 \begin{equation}\label{eq:A}
 {\mathscr A}_{\lambda\sigma}=\delta_{\lambda\sigma}\sum_{\alpha\beta}{\mathscr U}^\dagger_{\lambda \beta}G_{\alpha\beta}(0){\mathscr U}_{\alpha\lambda}.
\end{equation}
% comprised of the diagonal elements of the matrix ${\mathscr U}^\dagger \left({G^0}\right)^T{\mathscr U}$ and the superscript $T$ indicates matrix transpose.
  Two conditions follow immediately: (i) $G^T$ and $J$ must have the same eigenvectors since the same unitary operator $\mathscr U$ diagonalizes both of them, and (ii) the eigenvalues $g_\sigma$ of $G^T$ are the diagonal elements of $ {\mathscr A}$ and related to the initial correlation functions as follows:
 \begin{equation}\label{eq:lambda}
g_\sigma=\sum_\alpha |{\mathscr U}_{\alpha \sigma}|^2f(\mu_\alpha)\equiv \sum_\alpha W_{ \sigma\alpha}f(\mu_\alpha),
\end{equation}
where the matrix $W$ of transition probabilities is defined through \begin{equation}\label{eq:W}
 W_{\sigma\alpha }=|{\mathscr U}_{\alpha \sigma}|^2.
 \end{equation}
  We have inserted Eq.~\eqref{eq:corr0} into Eq.~\eqref{eq:A} to obtain the condition~\eqref{eq:lambda}. The matrix $W$ is a doubly stochastic (it has non-negative elements and the sum of every row and every column is equal to unity), which implies the sum rule $\sum_\alpha g_\alpha=\sum_\alpha f(\mu_\alpha)$. Given a matrix $G$ of desired steady-state correlators, we can find $g_\sigma$ and $\mathscr U$ by diagonalizing $G$. The matrix $W$ can then be explicitly constructed. For a given initial inverse temperature $\bbbeta$, we can then write the chemical potentials in the initial Hamiltonian as
 \begin{equation}\label{eq:mu}
\mu_j={1\over \bbbeta}\ln\left({1\over\sum_\sigma W^{-1}_{j\sigma}g_\sigma}-1\right).
\end{equation}

The columns of the matrix ${\mathscr U}=\left(|g_1\rangle, |g_2\rangle, \dots\right)$ are eigenvectors $|g_\alpha\rangle$ of $G=\sum_\alpha g_\alpha|g_\alpha\rangle \langle g_\alpha|$. The matrix $J$ can then be any linear combination 
 \begin{equation}\label{eq:J}
J=\sum_\alpha d_\alpha |g_\alpha\rangle \langle g_\alpha| \end{equation}
as long as the eigenvalues $\{d_\alpha\}$ are nondegenerate.

It may appear that the Hermitian matrix $G$ can be completely arbitrary, and we can create any correlation functions we choose. Hermiticity is not the only requirement of a physical matrix of Green's functions, however. For example, all diagonal elements are the occupation number of a site and must be between 0 and 1, so creating states with nonphysical occupation numbers should be impossible. Furthermore, the method fails for many physically allowed Green's functions, as discussed below.

 First, the matrix $W$ must be invertible, which is not guaranteed. Indeed we have found that $W$ tends to be singular for uniform systems, and some disorder in the occupation numbers is necessary for its invertibility. Second, the Fermi-Dirac functions are bounded as $0\leqslant f(\mu_\alpha)\leqslant1$, which is equivalent to the condition that the chemical potentials written in Eq.~\eqref{eq:mu} are real numbers. Therefore, the eigenvalues and eigenvectors of $G$ must be consistent with the following constraint: 
\begin{equation}\label{eq:constraint}
0 \leqslant\sum_\sigma W^{-1}_{j\sigma}g_\sigma\leqslant 1
\end{equation}
for all $j$. 

%
%
%Since the matrix $W$ is doubly stochastic, each $g_\sigma$ is a weighted average of Fermi-Dirac functions, and we must have $0\leqslant\lambda_\sigma\leqslant1$ for all $\sigma$.

\begin{figure}
\includegraphics*[width=1.0\columnwidth]{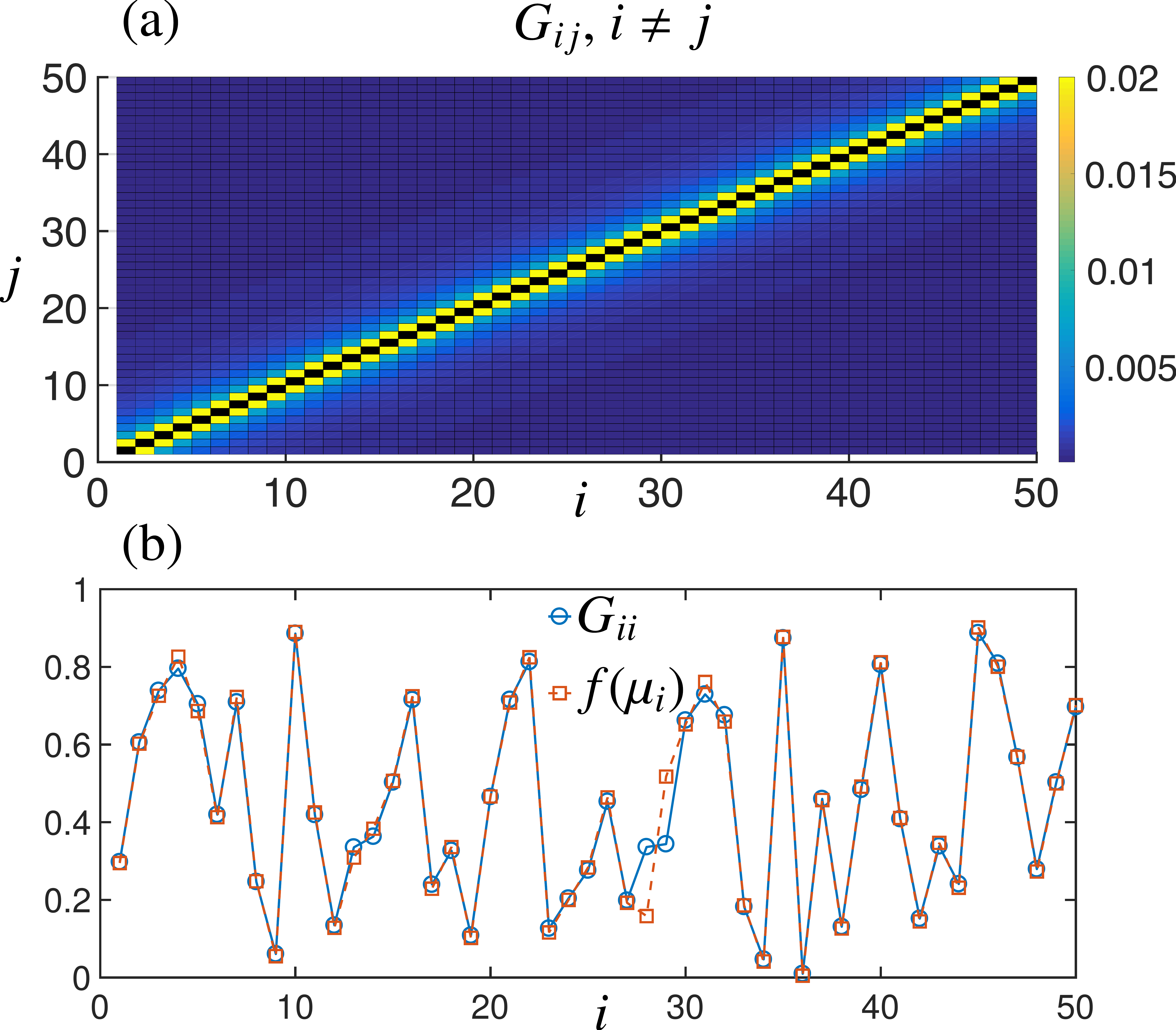}
\caption{(a) An example of the target Green's function $G$ with power-law off-diagonal elements for a system of $N=50$ sites, with $c=0.02$, and $\nu=1.5$ [see Eq.~\eqref{eq:exp}]. Only the off-diagonal elements are illustrated in panel (a). (b) The diagonal disordered elements of $G$, i.e., the target steady-state occupation numbers, and the corresponding initial occupation numbers $f(\mu_i)$ form Eq. \eqref{eq:mu}. 
\label{fig:1}}
\end{figure}
The constraint of Eq.~\eqref{eq:constraint} is not too restrictive if we are only interested in a pattern of correlations between different sites (off-diagonal elements of $G$) without simultaneously specifying on-site occupation numbers (diagonal elements of $G$). This observation is because we can change the eigenvalues of $G$ by adding a multiple of the identity matrix to $G$, which only changes the occupation numbers, or multiplying $G$ by a constant, which only rescales the Green's functions, without changing their overall pattern. More specifically, suppose for some matrix $G$,  $a=\min \left(\sum_\sigma W^{-1}_{j\sigma}g_\sigma \right)<0$ and $b=\max \left(\sum_\sigma W^{-1}_{j\sigma}g_\sigma \right) >1$. Transforming $G\to{1\over b-a}(G-aI)$ by a shift to the diagonal elements and a rescaling guarantees the constraint~\eqref{eq:constraint} is satisfied. Here, $I$ is the identity matrix. 

While the initial chemical potentials $\mu_j$ are uniquely determined by $G$ and $\bbbeta$ as shown in Eq.~\eqref{eq:mu}, we have many choices for $J_{ij}$. Essentially, the only constraint on $J$ is through its eigenvectors, and we can choose the eigenvalues at will as long as they are nondegenerate. The absence of degeneracy is essential for our method. For example, in the extreme case where all eigenvalues $d_\alpha$ are degenerate, the matrix $J$ becomes proportional to the identity matrix, which will not change the initial Green's functions. The hopping matrix $J$ is also entirely local in the limit of all degenerate eigenvalues. There is indeed competition between locality and equilibration time. To speed up the emergence of the steady state, we spread out these eigenvalues and make the gaps $d_\alpha-d_\sigma$ as large as possible. 

The timescale for the equilibration goes as
 \begin{equation}t_{\rm eq}={1\over \left[\min (d_\alpha-d_\sigma)\right]^2},
 \end{equation} according to Eq. \eqref{eq:master_sol}. Of course, we can  arbitrarily shorten the timescale if we allow the energy scale of the Hamiltonian to run to infinity. To find an optimal $J$ with fixed energy scale, we set the spectral norm of $J$, i.e., the square root of the largest eigenvalue of the matrix $J^\dagger J$  to unity, $||J||=1$, which implies $ -1\leqslant d_\alpha \leqslant 1$.
To minimize $t_{\rm eq}$, we can make all the gaps identical by choosing
 \begin{equation}\label{eq:d_order}
d_\alpha=-1+2{\alpha-1\over N-1},\quad \alpha=1,\dots, N.
 \end{equation}
While using all permutations of the above values of $d_\alpha$ give the same equilibration time, we have found that the smoothest site dependence in matrix $J$ corresponds to the case where eigenvectors of $G$ are ordered according to their eigenvalues, i.e., $
g_\alpha\leqslant g_{\alpha+1}$.

\section{Engineering power-law two-point functions \label{sec:power}}

We now consider an example of creating off-diagonal Green's functions
\begin{equation}\label{eq:exp}
G_{ij}={c\over |i-j|^\nu}, \quad i\neq j,
\end{equation}
for an arbitrary exponent $\nu$. An example of a target $G_{ij}$ for $i\neq j$ is shown in Fig.~\ref{fig:1}(a) for $\nu=1.5$. We found that requiring a uniform average density gives a singular $W$ matrix. Thus, to make the matrix $W$ invertible, we choose random values drawn from a uniform distribution for the diagonal elements $G_{ii}$ (note that $G_{ii}$ is the expectation value of the occupation number of fermionic mode and must be between 0 and 1 for any physical system). Our results on the structure of the Hamiltonian $J$, the emergence of the steady-state correlations and the behavior of the four-point functions are not sensitive to the realization of diagonal disorder.

\begin{figure}
\includegraphics*[width=1.0\columnwidth]{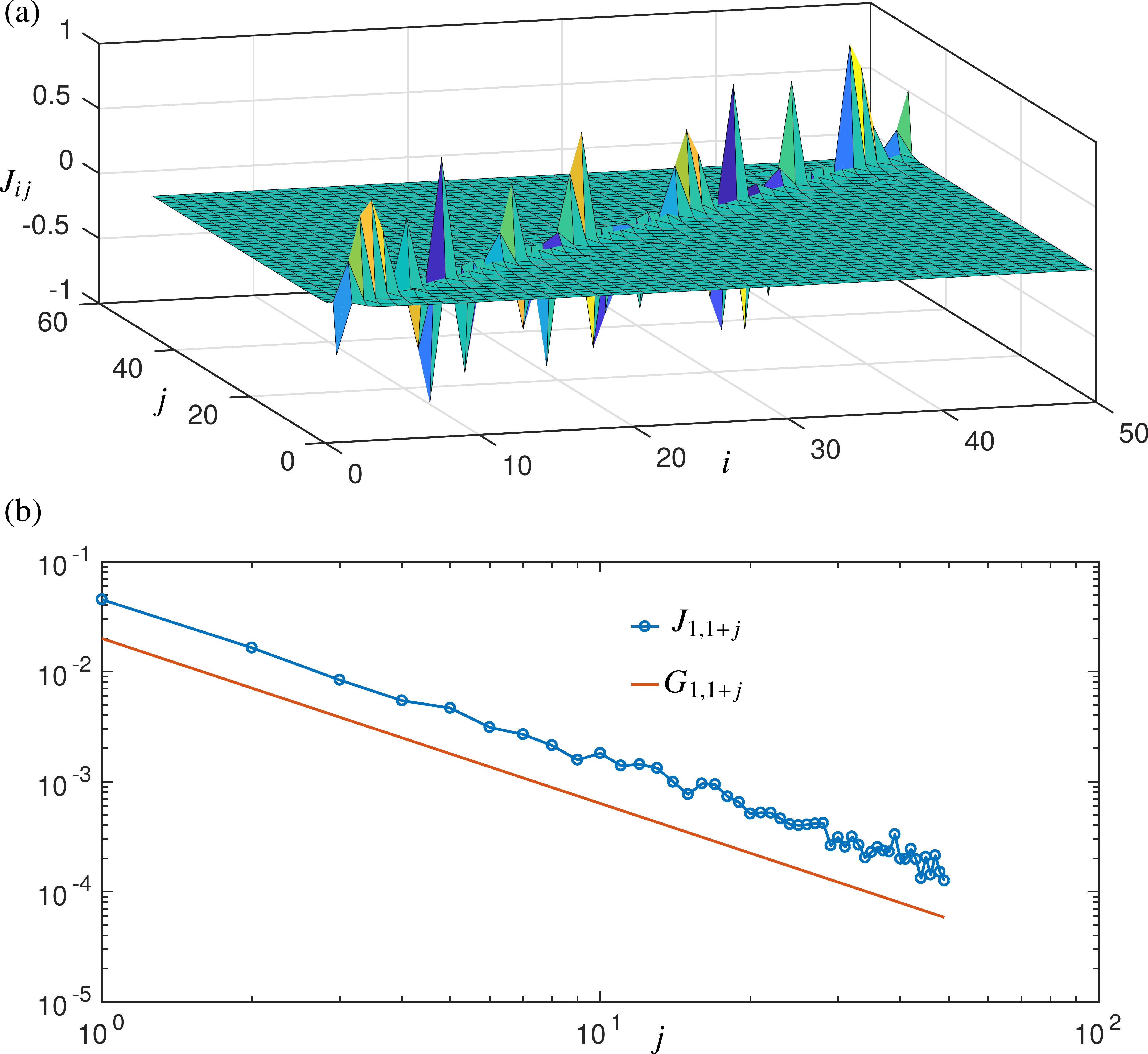}
\caption{(a) The elements of the drive Hamiltonian $J$. The largest elements are local in the vicinity of the diagonal, but $J$ exhibits nonlocal power-law decay in off-diagonal elements. (b) The elements of $J$ decay with the same $\nu$ exponent (in this example $\nu=1.5$ as the two-point function).
\label{fig:2}}
\end{figure}
\begin{figure}
\includegraphics*[width=1.0\columnwidth]{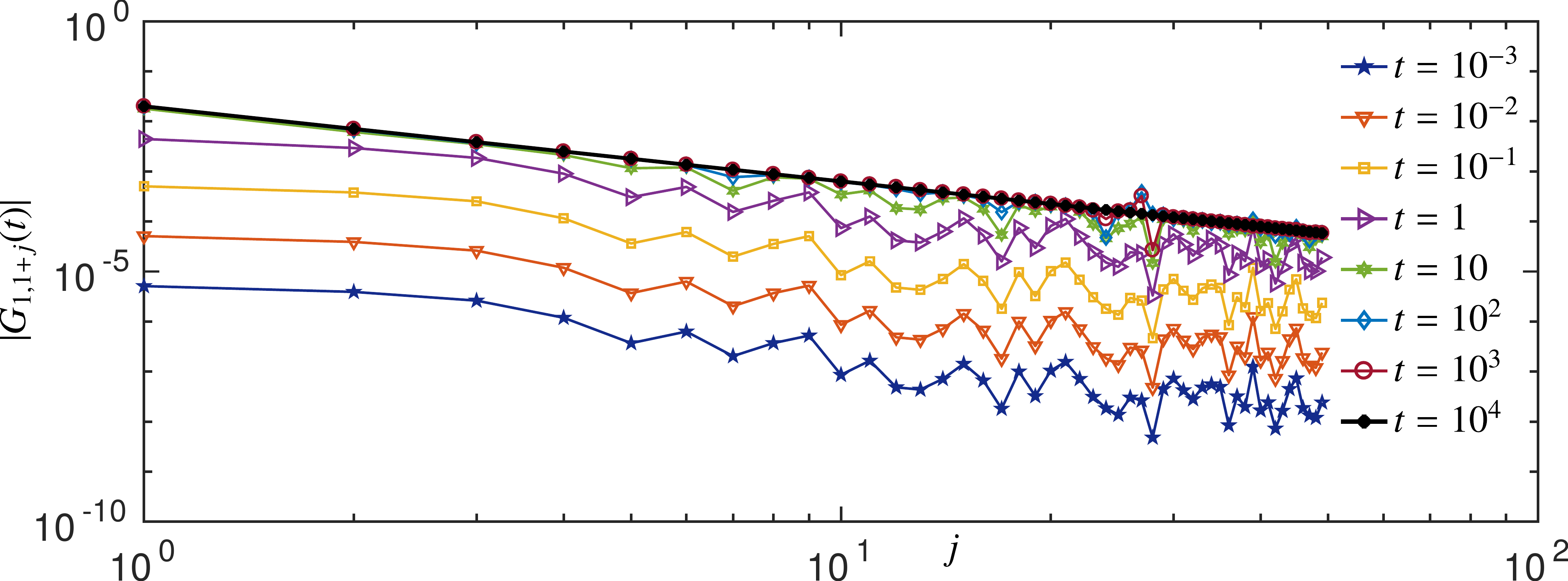}
\caption{The time-dependent approach of the two-point function to the engineered steady-state values.
\label{fig:3}}
\end{figure}
To apply our method, we find the eigenvalues and eigenvector of $G$, which in turn yield the initial chemical potentials according to Eq.~\eqref{eq:mu}. As long as the initial Fermi-Dirac functions $f(\mu_i)$, which only depend on the eigenvalues and eigenvectors of $G$, are between 0 and 1, we can find real chemical potential for any $\bbbeta$. As an example, we show in Fig.~\ref{fig:1}(b) the $f(\mu_i)$ that yield the Green's functions shown in Fig.~\ref{fig:1}, together with the corresponding $G_{ii}$. The initial occupation numbers $f(\mu_i)$ are close to the target occupation numbers $G_{ii}$ but not identical. The smallest difference for $i=47$ is $0.13\%$, while the largest difference for $i=28$ is $53\%$. We note that while the initial occupation numbers are redistributed in the final steady state, the total charge is conserved, and we have $\sum_i G_{ii}=\sum_i f(\mu_i)$.

In Fig.~\ref{fig:2}, we show the elements of the $J$ matrix of Eq.~\eqref{eq:J} obtained by eigenvalues \eqref{eq:d_order}. We find that the elements of $J$ exhibit similar critical correlations as the correlations we seek to create. We examined many different permutations of the eigenvalues. While the $J$ can become highly disordered, it appears that the correlation structure does not change from critical as long as we keep the eigenvalues separated for optimal equilibration.

It is illuminating to examine the time dependence of the two-point function as it approaches the engineered critical correlations. As shown in Fig.~\ref{fig:3}, they start from zero and quickly acquire a noisy pattern aligned with the overall power-law dependence. The evolution continuously shifts the correlators up in a log-log plot while reducing the noise until it converges to the target disorder-free power law. The exponent of the critical correlation emerges much faster than the quantitative emergence of those correlation functions.

\section{Density-density correlations in the steady state \label{sec:NN}} 
\begin{figure}
\includegraphics*[width=.89\columnwidth]{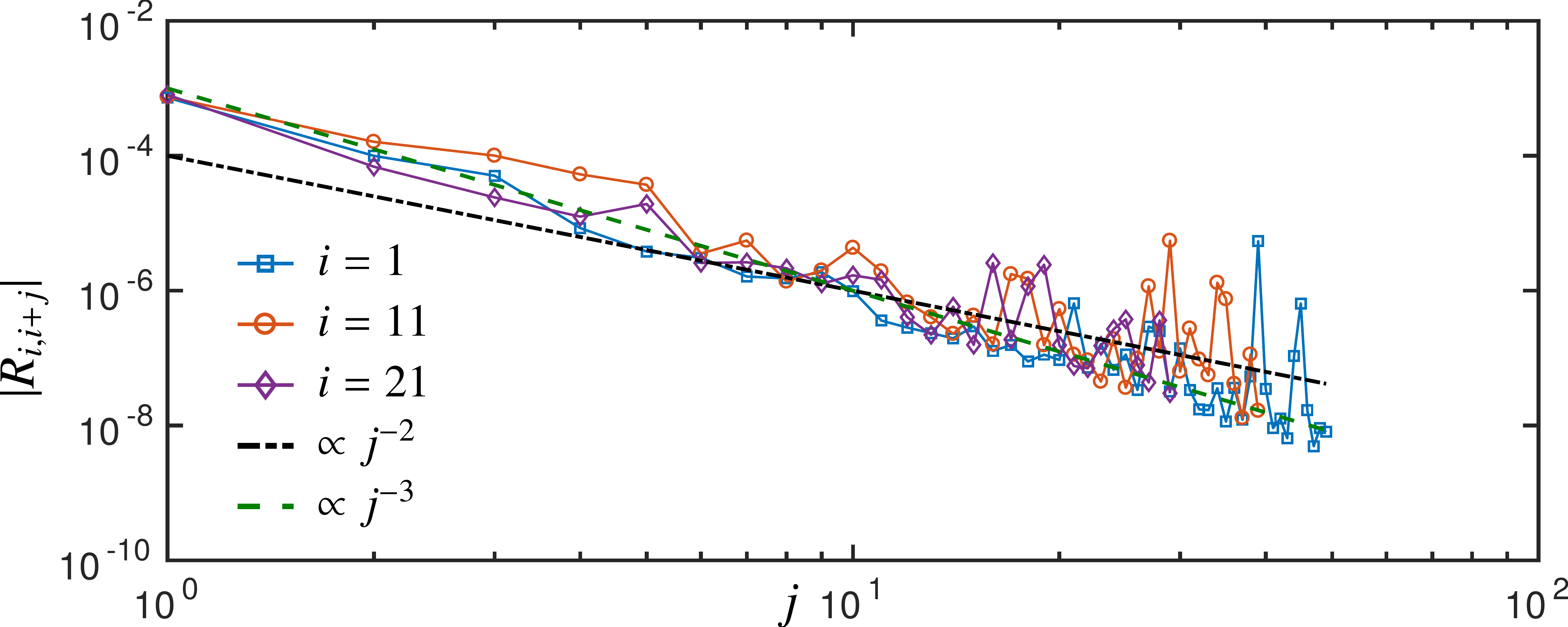}
\caption{The density-density correlation functions in the steady state with a $j^{-1.5}$ Green's function. Despite strong disorder, the correlation decays with a trendline of $j^{-3}$.
\label{fig:4}}
\end{figure}

To understand the nature of the steady state we have generated, we compute the density-density correlation functions, as the two-point function does not fully characterize the steady state. A widely studied class of correlation functions is the connected density-density correlator
\begin{equation}
R_{ij}(t)=\langle \hat n_i \hat n_j\rangle-\langle \hat n_i\rangle\langle \hat n_j\rangle,
\end{equation} 
We treat the two terms in the above expression separately. To find $\langle \hat n_i \hat n_j\rangle$, we use the formalism of Sec.~\ref{sec:formalism} to obtain the long-time limit of the rank-4 tensor $\mathscr R$ for the density-density operator. We can then obtain the time dependence as well as the long-time limit of the $\langle \hat n_i \hat n_j\rangle$ correlator in terms of the initial four-point-function expectation values $\langle c^\dagger_\alpha c_\beta c^\dagger_\gamma c_\delta\rangle_{(0)}$ at $t=0$, which using the Wick's theorem is given by
\begin{equation}\label{eq:4_init}
\langle c^\dagger_\alpha c_\beta c^\dagger_\gamma c_\delta\rangle_{(0)}=\delta_{\beta \gamma}\delta_{\alpha\delta}[1-f(\mu_\beta)]f(\mu_\alpha)+\delta_{\alpha\beta}\delta_{\gamma\delta}f(\mu_\alpha)f(\mu_\gamma).
\end{equation}
\begin{widetext}
%We begin with the time-dependent expression
%\begin{equation}
%\langle \hat n_i \hat n_j\rangle_{(t)}=\sum_{\alpha\beta\gamma\delta,abcd}{\mathscr U}_{\alpha a}{\mathscr U}_{ib}{\mathscr U}_{\gamma c}{\mathscr U}_{jd}e^{-{1\over 2}\left({\cal D}_a+{\cal D}_c-{\cal D}_b-{\cal D}_d\right)^2t}{\mathscr U}^\dagger_{d\delta}{\mathscr U}^\dagger_{cj}{\mathscr U}^\dagger_{b\beta}{\mathscr U}^\dagger_{ai}\langle c^\dagger_\alpha c_\beta c^\dagger_\gamma c_\delta\rangle_{(0)}
%\end{equation}
%whose 
The long-time limit for $t\to\infty$ follows from Eq. \eqref{eq:4_long} as shown below
\begin{equation}
\langle \hat n_i \hat n_j\rangle_{(\infty)}=\sum_{\alpha\beta\gamma\delta}\left[\sum_{ac}
{\mathscr U}_{\alpha a}{\mathscr U}_{\gamma c}{\mathscr U}^\dagger_{cj}{\mathscr U}^\dagger_{ai}
\left({\mathscr U}_{ia}{\mathscr U}_{jc}{\mathscr U}^\dagger_{c\delta}{\mathscr U}^\dagger_{a\beta}+{\mathscr U}_{ic}{\mathscr U}_{ja}{\mathscr U}^\dagger_{a\delta}{\mathscr U}^\dagger_{c\beta}\right)-
\sum_a{\mathscr U}_{\alpha a}{\mathscr U}_{\gamma a}{\mathscr U}^\dagger_{aj}{\mathscr U}^\dagger_{ai}
{\mathscr U}_{ia}{\mathscr U}_{ja}{\mathscr U}^\dagger_{a\delta}{\mathscr U}^\dagger_{a\beta}\right]\langle c^\dagger_\alpha c_\beta c^\dagger_\gamma c_\delta\rangle_{(0)}.
\end{equation}
It is convenient to define column vectors $F=(f(\mu_1), f(\mu_2), \dots ,f(\mu_N))$, $K=(1-f(\mu_1), 1-f(\mu_2), \dots ,1-f(\mu_N))$, and matrices $A^{ij}={\mathscr U}{\rm diag}({\mathscr U}_{i,1}{\mathscr U}^\dagger_{1,j}, {\mathscr U}_{i,2}{\mathscr U}^\dagger_{2,j}, \dots ,{\mathscr U}_{i,N}{\mathscr U}^\dagger_{N,j}){\mathscr U}^\dagger$,
% such that
%\[
%F=(f(\mu_1), f(\mu_2), ...,f(\mu_N)), \qquad K=(1-f(\mu_1), 1-f(\mu_2), ...,1-f(\mu_N)), \qquad A^{ij}={\mathscr U}{\rm diag}({\mathscr U}_{i,1}{\mathscr U}^\dagger_{1,j}, {\mathscr U}_{i,2}{\mathscr U}^\dagger_{2,j},...,{\mathscr U}_{i,N}{\mathscr U}^\dagger_{N,j}){\mathscr U}^\dagger,
%\]
where $f(\mu_\alpha)$ is defined in Eq.~\eqref{eq:corr0}. Inserting Eq. \eqref{eq:4_init} into the above expression, upon some matrix algebra, leads to
\begin{equation}
\begin{split}
\langle \hat n_i \hat n_j\rangle_{(\infty)}=\sum_{\alpha \beta}\Big\{\big[A^{ii}_{\alpha\beta}A^{jj}_{\beta\alpha}+A^{ji}_{\alpha\alpha}A^{ij}_{\beta\beta}\big] F_\alpha K_\beta+
\big[A^{ii}_{\alpha\alpha}A^{jj}_{\beta\beta}+A^{ji}_{\alpha\beta}A^{ij}_{\beta\alpha}\big] F_\alpha F_\beta-W_{\beta, i}W_{\beta \alpha}F_{\alpha}W_{\beta,j},\Big\}
\end{split}
\end{equation}
\end{widetext}
where the $W$ matrices are defined in Eq.~\eqref{eq:W}. The subtraction of the background $\langle \hat n_i\rangle\langle \hat n_j\rangle$ is subtle. It might appear that each $\langle \hat n_i\rangle$ in the long-time limit is simply a diagonal element of $G$. However, that is only correct if we first take the noise average of each expectation value of density and multiply the noise-averaged results. It is more natural to take the quantum averages of density for various realizations of noise first, multiply them, and then take the noise average,  keeping the noise averaging consistently in the last step. The long-time limit can be obtained by time averaging the product of expectation values upon a quantum quench to Hamiltonian $V$. In this approach, the background term yields two terms that are identical to terms in $\langle \hat n_i \hat n_j\rangle_{(\infty)}$. We find
\[
\langle \hat n_i\rangle\langle \hat n_j\rangle=\sum_{\alpha \beta}\
\big[A^{ii}_{\alpha\alpha}A^{jj}_{\beta\beta}+A^{ji}_{\alpha\beta}A^{ij}_{\beta\alpha}\big] F_\alpha F_\beta-\sum_\alpha W_{i, \alpha} W_{j, \alpha}\left(\tilde F_{\alpha\alpha}\right)^2,
\]
where the matrix $\tilde F={\mathscr U}^\dagger{\rm diag}(F_1, \dots, F_N){\mathscr U}$. Using these expressions we calculated the long-time limit of $R_{ij}$ for the same drive that produces power-law two-point functions of Eq.~\eqref{eq:exp}. As shown in Fig.~\ref{fig:4}, while the long-time limit of $G$ has no disorder in the off-diagonal terms, the density-density correlation functions are highly disordered. The minima in $|R_{i,j}|$ decay as $|j-i|^{-3}$ for $\nu=1.5$. It seems that the steady states emerging upon engineering clean $x^{-\nu}$ two-point functions have highly disordered density-density four-point functions with a trendline scaling as $x^{-2\nu}$. 

%$G_{ij}^0={1\over Z}{\sum_{\{n_k=0,1\}}\langle n_1\dots n_N|c^\dagger_ic_j|n_1\dots n_N\rangle e^{-\bbbeta(\mu_1n_1+\dots \mu_Nn_N)}  }$, which leads to
\section{Conclusions\label{sec:conclusions}}
In this paper, we extended the Heisenberg-picture formalism of stochastic driving to the case of fermionic four-point functions. We then proposed a scheme of reverse-engineering the drive to generate customized fermionic two-point functions from a completely uncorrelated initial thermal state. Finally, we examined the case of generating critical correlations and found that the steady state exhibits clean engineered two-point functions and highly disordered density-density correlation functions. Such steady states with custom-ordered correlations provide novel examples of drive engineering.
\section{Acknowledgments}
I thank Michael Kolodrubetz and Zohar Nussinov for helpful discussions. 
This work was supported by NSF Award No. DMR-1945395. I thank the Kavli Institute for Theoretical Physics, which is supported by
the NSF Grant No. PHY-1748958, for hospitality. I also thank the Aspen Center for Physics, which is supported by NSF Grant No. PHY-1607611, for hospitality.
\bibliographystyle{apsrev4-1}
\bibliography{references}

\end{document}